\documentclass[10pt]{article} 
\usepackage[utf8]{inputenc} 
\usepackage{setspace} 
\usepackage{graphicx}
\usepackage{geometry} 
\usepackage{epsf}
\usepackage{enumerate}
\usepackage{natbib}
\usepackage{url} 

\usepackage{subcaption}
\usepackage{listings}
\usepackage{textcomp}
\usepackage{hyperref}
\usepackage{amssymb,amsmath}
\usepackage{amsmath,amsthm,mathtools}
\usepackage{bbm}
\usepackage{units}
\usepackage{tikz}
\usepackage{qtree}
\usetikzlibrary{trees}
\usepackage{tikz-qtree,tikz-qtree-compat}
\tikzset{hide on/.code={\only<#1>{\color{white}}}}
\usepackage{algorithm2e}
\makeatletter
\renewcommand{\@algocf@capt@plain}{above}
\makeatother

\newcommand\ceil[1]{\lceil#1\rceil}

\usepackage{multirow}

\newtheorem{theorem}{Theorem}
\theoremstyle{plain}

\begin{document}

  \title{\bf Tree-based Particle Smoothing Algorithms in a Hidden Markov Model}
  \author{Dong Ding \hspace{.2cm}\\
    Department of Mathematics, Imperial College London\\
    Axel Gandy \\
    Department of Mathematics, Imperial College London}
  \maketitle

\begin{abstract}
We provide a new strategy built on the divide-and-conquer approach by Lindsten et al. (2017, Journal of Computational and Graphical Statistics) to investigate the smoothing problem in a hidden Markov model. We employ this approach to decompose a hidden Markov model into sub-models with intermediate target distributions based on an auxiliary binary tree structure and produce independent samples from the sub-models at the leaf nodes towards the original model of interest at the root. 
We review the target distribution in the sub-models suggested by Lindsten et al. and propose two new classes of target distributions, which are the estimates of the (joint) filtering distributions and the (joint) smoothing distributions. The first proposed type is straightforwardly constructible by running a filtering algorithm in advance.  The algorithm using the second type of target distributions has an advantage of roughly retaining the marginals of all random variables invariant at all levels of the tree at the cost of approximating the marginal smoothing distributions in advance. We further propose parametric and non-parametric ways of constructing these target distributions using pre-generated Monte Carlo samples.
We show empirically the algorithms with the proposed intermediate target distributions give stable and comparable results as the conventional smoothing methods in a linear Gaussian model and a non-linear model. 
\end{abstract}

\noindent%
{\it Keywords: Algorithms; Bayesian methods; Monte Carlo simulations; Particle filters}  
\vfill

\newpage

\section{Introduction}
\label{intro}
 
A hidden Markov model (HMM) is a discrete-time stochastic process
$\{X_{t}, Y_{t}\}_{t \geq 0}$ where $\{X_{t}\}_{t \geq 0}$ is an
unobserved Markov chain.  We only have access to $\{Y_{t}\}$ whose
distribution depends on $\{X_{t}\}$.  We make the following
assumptions in the entire article:
The densities of the initial state $X_{0}$, the transition density $X_{t+1}$ given $X_{t} = x_{t}$ and the emission density $Y_{t}$ given $X_{t} = x_{t}$ taken with respect to some dominating measure exist and are denoted as follows:
\begin{alignat*}{2}
X_{0} &\sim p_{0}(x_{0}) &&\\
X_{t+1}| \{ X_{t} = x_{t} \} &\sim p(x_{t+1} | x_{t}) && \text{~~for }  t = 0, \ldots, T-1,\\
Y_{t} | \{X_{t} = x_{t}\} &\sim p(y_{t}|x_{t}) && \text{~~for }  t = 0, \ldots, T,
\end{alignat*}
where $T$ is the final time step of the process. 

We are interested in the (marginal) smoothing distributions
$\{p(x_{t}|y_{0:T})\}_{t = 0, \ldots, T}$ or the joint smoothing
distribution $p(x_{0:T}|y_{0:T})$ where $x_{0:T}$ and $y_{0:T}$
are abbreviations of $(x_{0}, \ldots, x_{T})$ and
$(y_{0}, \ldots, y_{T})$, respectively. Exact solutions are available
for linear Gaussian HMM using a Rauch--Tung--Striebel smoother (RTSs)
\citep{rauch1965maximum} and in a HMM with finite-space Markov chains
\citep{baum1966statistical}. In most other cases, the smoothing
distributions are not analytically tractable.

A large body of work uses Monte Carlo methods to approximate the
smoothing distributions $\{p(x_{t}|y_{0:T})\}_{t = 0, \ldots, T}$ or
the joint smoothing distribution $p(x_{0:T}|y_{0:T})$.  Sequential
Monte Carlo (SMC) methods \citep{de2001sequential} are commonly used
to sequentially update the filtering distributions
$\{p(x_{t}|y_{0:t})\}_{t = 0, \ldots, T}$.  SMC can in principle be
used to estimate the joint smoothing density $p(x_{0:T}|y_{0:T})$ by
updating the entire history of the random samples in each resampling
step. However, the performance can be poor, as path degeneracy will
occur in many settings \citep{arulampalam2002tutorial}. Advanced
sequential Monte Carlo methods with desirable theoretical and
practical results have been developed in recent years including
sequential Quasi-Monte Carlo (SQMC) \citep{gerber2015sequential},
divide-and-conquer sequential Monte Carlo (D\&C SMC)
\citep{lindsten2017divide}, multilevel sequential Monte Carlo (MSMC)
\citep{beskos2017multilevel} and variational sequential Monte Carlo
(VSMC) \citep{naesseth2017variational}.

Other smoothing algorithms have been suggested
previously. \citet{doucet2000sequential} develop the forward filtering
backward smoothing algorithm (FFBSm) for sampling from 
$\{p(x_{t}|y_{0:T})\}_{t = 0, \ldots, T}$ based on the formula
proposed by \citet{kitagawa1987non}. \citet{godsill2004monte} propose
the forward filtering backward simulation algorithm (FFBSi) which
generates samples from the joint smoothing distribution
$p(x_{0:T}|y_{0:T})$. \citet{briers2010smoothing} propose a two-filter
smoother (TFS) which employs a standard forward particle filter and a
backward information filter to sample from
$\{p(x_{t}|y_{0:T})\}_{t = 0, \ldots, T}$.  Typically, these
algorithms have quadratic complexities in $N$ for generating $N$
samples.  \citet{fearnhead2010sequential} and \citet{klaas2006fast}
propose two smoothing algorithms with lower computational complexity,
but their methods do not provide unbiased estimates.

In this article, we suggest using the divide-and-conquer sequential
Monte Carlo (D\&C SMC) \citep{lindsten2017divide} approach to address
the smoothing problem. The D\&C SMC algorithm performs statistical
inferences in probabilistic graphical models. It splits the random
variables of the target distribution into multiple levels of disjoint
sets based upon an auxiliary tree $\mathcal{T}$. An intermediate
target distribution needs to be assigned to each set of random
variables yielding sub-models for each non-leaf node.  The choice of
these intermediate target distributions is key for a good overall
performance of the algorithm. By sampling independently from the leaf
nodes and gradually propagating, merging and resampling from the leaf
nodes to the root, the D\&C SMC algorithm eventually produces samples
from the target distribution. The merging step involves importance
sampling.

Using the idea of D\&C SMC, we aim to estimate the joint smoothing
distribution $p(x_{0:T}|y_{0:T})$ and thus call the algorithm:
`tree-based particle smoothing algorithm' (TPS). The key differences
between TPS and other smoothing algorithms lie in its non-sequential
and more adaptive merging step of the samples.

Our main contribution is the proposition and investigation of three
classes of intermediate target distributions to be used in the
algorithm.  We denote a leaf node corresponding to a single random
variable $X_{j}$ by $\mathcal{T}_{j} \in \mathcal{T}$ and a non-leaf
node corresponding to the random variables $X_{j:l}$ by
$\mathcal{T}_{j:l} \in \mathcal{T} (j < l)$.

The first class advised by \cite{lindsten2017divide} has the density proportional to the product
of all transition and emission densities associated to the target variable $X_{j}$ (resp. $X_{j:l}$) in the sub-model. This is equivalent to the unnormalised likelihood of a new HMM starting at time $j$ (resp. from time $j$ to $l$) given the observations of the same time interval with an uninformative prior of $X_{j}$ if $j \neq 0$.

The second class uses an estimate of the filtering
distribution $p(x_{j} | y_{0:j})$ at $\mathcal{T}_{j} \in \mathcal{T}$
and an estimate of the joint filtering distribution
$p(x_{j:l} | y_{0:l})$ at $\mathcal{T}_{j:l} \in \mathcal{T}$. Working
with this estimate involves tuning a preliminary particle filter.

The third class uses estimates of the marginal smoothing distribution
$p(x_{j} | y_{0:T})$ at $\mathcal{T}_{j} \in \mathcal{T}$ and of the
joint smoothing distribution $p(x_{j:l} | y_{0:T})$ at
$\mathcal{T}_{j:l} \in \mathcal{T}$. We will see that this class of
immediate distributions is optimal in a certain sense.  Furthermore,
under this construction, we approximately retain the marginal
distribution of all single random variables $\{X_{j}\}_{j = 0}^{T}$
invariant as the marginal smoothing distributions
$\{ p(x_{j} | y_{0:T}) \}_{j = 0}^{T}$ at every level of the tree. The
price of implementing TPS using the second class of intermediate
target distributions relies on both the estimates of the filtering and
the (marginal) smoothing distributions, but not necessarily the joint
smoothing distribution.  We then propose some parametric and
non-parametric approaches to construct these intermediate
distributions based on the pre-generated Monte Carlo samples
considering both efficiency and accuracy.


The article is structured as follows. We first describe the divide-and-conquer approach for particle smoothing in Section \ref{TPS_intro}. We discuss the intermediate target distributions and the constructions of the initial sampling distributions at the leaf nodes in Section \ref{target_distr}. In Section \ref{simulation}, we conduct simulation studies in a linear Gaussian and non-linear non-Gaussian HMM to compare TPS with other smoothing algorithms. The article finishes with a discussion in Section \ref{conclusion}.

\section{Tree-based Particle Smoothing Algorithm (TPS)} 
\label{TPS_intro}
This section outlines an algorithm we call `tree-based particle smoothing algorithm' (TPS). 
\cite{lindsten2017divide} describe the construction of an auxiliary tree for  general probabilistic graphical models.
We demonstrate a unique construction of an auxiliary binary tree from a HMM bearing intermediate target distributions specified at each node. We then illustrate the sampling procedure for the target distributions at the nodes. We present an algorithm which can be applied recursively from the leaf nodes towards the root and yet generate the target samples.

\subsection{Construction of an auxiliary tree}
\label{Aux_tree}
TPS splits a HMM into sub-models based upon a binary tree
decomposition. It first divides the random variables $X_{0:T}$ into
two disjoint subsets and recursively apply binary splits to the
resulting two subsets until the resulting subset consists of only a
single random variable. Each generated subset corresponds to a tree
node and is assigned an intermediate target distribution. The root
characterises the complete model with the target distribution
$p(x_{0:T}|y_{0:T})$.  Initial samples are generated at the leaf
nodes, independent between leaves. Theses samples are recursively
merged using importance sampling until the root of the tree is
reached.

\begin{figure}[tbp]
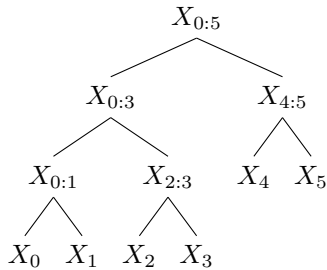

\centering
\Tree    [. $X_{0:5}$ [.   $X_{0:3}$   [.  $X_{0:1}$ [.   $X_{0}$ ] [.  $X_{1}$  ]] [.  $X_{2:3}$   [.  $X_{2}$  ] [. $X_{3}$ ]]   ]   [.   $X_{4:5}$   [.   $X_{4}$ ]   [.    $X_{5}$  ]  ] ]  
\caption{An auxiliary binary tree consisting of random variables when $T = 5$}
\label{RVoT}
\end{figure}

We propose one intuitive way of implementing the binary splits which
ensures that the left subtree is always a complete binary tree and
contains at least as many nodes as the right subtree. We split a non-leaf
node with the  variables $X_{j:l}$ where
$0 \leq j < l \leq T$, into  two
children $\mathcal{T}_{j:k-1}$ and $\mathcal{T}_{k:l}$ with the random
variables $X_{j:k-1}$ and $X_{k:l}$, where

\begin{eqnarray}
\label{cut_point}
k = j + 2^{p},
\end{eqnarray}
and $p = \ceil {\frac{\log(l-j+1)}{\log 2}} - 1.$  The auxiliary tree
when $T = 5$ is shown in Figure \ref{RVoT}.

This construction has several advantages: The random variables within
each node have consecutive time indices. The left subtree
is also a complete binary tree of $2^{\ceil {\frac{\log(T+1)}{\log 2}}}$
leave nodes. 
$\{y_{T+1}, y_{T+2}, \ldots\}$ become available, as samples from the
complete subtree would not need to be updated.

Moreover, the tree has a height of
$\big(\ceil {\frac{\log(T+1)}{\log 2}} + 1\big)$ levels, which implies
a maximum number of $\ceil {\frac{\log(T+1)}{\log 2}}$ updates of the
samples corresponding to a single random variable with different target
distributions at different levels of the tree. Usually, more updates potentially indicate more resampling steps, which may cause more serious degeneracy problems.
 In Figure \ref{RVoT},
the samples corresponding to $X_{0}, \ldots, X_{3}$ need to be updated
three times from the leave nodes and those of $X_{4}, X_{5}$ need to
be updated twice. When running a bootstrap particle filter to solve
the smoothing problem, the samples at time step $t = 0$ need to be
updated $T$ times and thus the maximum number of the updates become
$T$, which is no less than $ \ceil {\frac{\log(T+1)}{\log 2}}$. 

\cite{lindsten2017divide} also propose a general way of constructing the auxiliary tree in a self-similar model family, where a HMM belongs to. Their construction in the context of a HMM may not be identical to ours with no restriction on the choice of the cutting point.

\subsection{Sampling procedure in the sub-models of tree}
We describe the sampling approach from the target distribution at a
leaf and non-leaf node of the constructed binary tree $\mathcal{T}$
described in Section \ref{Aux_tree}. We denote a target density by
$f_{j}$ which can be straightforwardly sampled from at a leaf node
$\mathcal{T}_{j} \in \mathcal{T}$, a proper importance density by
$h_{j:l}$ and a target density by $f_{j:l}$ respectively at a non-root
tree node $\mathcal{T}_{j:l} \in \mathcal{T}$ where $0 < l - j <
T$. At the root,  the target density is always
$f_{0:T} = p(x_{0:T}|y_{0:T})$.

At a leaf node $\mathcal{T}_{j}$, we sample from $f_{j}$ directly.  At a non-root node $\mathcal{T}_{j:l}$, we employ an importance sampling step with the proposal $h_{j:l} = f_{j:k-1} f_{k:l}$ being the product of the target densities from the two children of $\mathcal{T}_{j:l}$.  
Practically, we merge the samples from $\mathcal{T}_{j:k-1}$ and $\mathcal{T}_{k:l}$ respectively and reweigh them. 


\begin{algorithm}[tbp]
\SetAlgoLined
\eIf{j = l}{ Simulate $x_{j}^{(i)} \sim f_{j}(x_{j})$   for $i = 1,2, \ldots, N$.
Return $\{  x_{l}^{(i)}, w_{l}^{(i)} = \frac{1}{N} \}_{i = 1}^{N}.$
}
{Let $p = \ceil{\frac{\log(l-j+1)}{\log 2}} - 1$ and $k = j+ 2^{p}$. \\
$\{ \tilde{x}_{j:k-1}^{(i)}, \tilde{w}_{j:k-1}^{(i)} \}_{i = 1}^{N} \leftarrow  \texttt{TS}(j,k-1)$ from $\mathcal{T}_{j:k-1}$ 
and $\{  \tilde{x}_{k:l}^{(i)}, \tilde{w}_{k:l}^{(i)} \}_{i = 1}^{N} \leftarrow  \texttt{TS}(k,l)$ from $\mathcal{T}_{k:l}$.\\
Denote the combined particles by $\{ \tilde{x}_{j:l}^{(i)}  =  (\tilde{x}_{j:k-1}^{({i})}, \tilde{x}_{k:l}^{({i})}), \tilde{w}_{j:l}^{(i)} = \tilde{w}_{j:k-1}^{({i})} \tilde{w}_{k:l}^{({i})} \}_{i = 1}^{N}$. \\
Update the unnormalised weights for $i = 1, \ldots, N$:
\begin{eqnarray}
\label{weight_formula}
\hat{w}^{(i)}_{j:l} =  \tilde{w}_{j:l}^{(i)} \frac{f_{j:l} (\tilde{x}_{j:l}^{(i)})}{ f_{j:k-1}  (  \tilde{x}^{({i})}_{j:k-1}) f_{k:l}(\tilde{x}^{({i})}_{k:l})}.
\end{eqnarray}
Resample $\big\{ \tilde{x}_{j:l}^{(i)},  \hat{w}^{(i)}_{j:l}  \big\}_{i = 1}^{N}$ to obtain the normalised weighted particles $\big\{ x_{j:l}^{(i)},  w^{(i)}_{j:l}  \big\}_{i = 1}^{N}$.
\\
Return $\big\{ x_{j:l}^{(i)},  w^{(i)}_{j:l}  \big\}_{i = 1}^{N}.$
}
\caption{Algorithm \texttt{TS}($j, l$) which generates weighted samples from the target $f_{j:l}$}
\label{RecursiveAlg}
\end{algorithm}

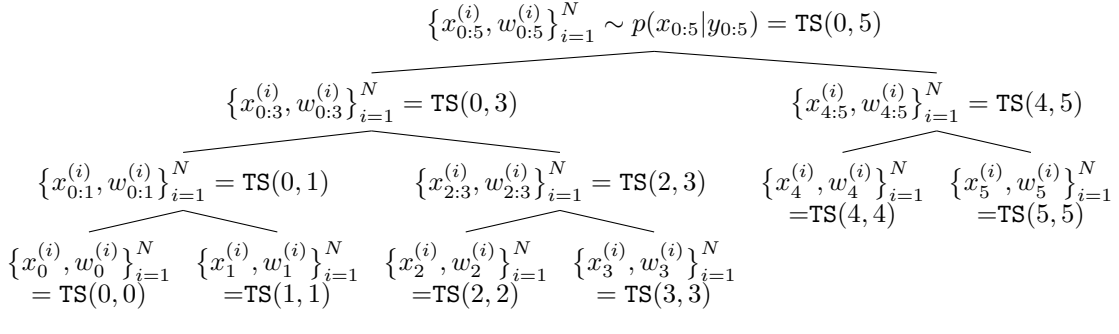
\begin{figure}[tbp]
\centering
\begin{tikzpicture}[scale=1]
\tikzset{every tree node/.style={align=center,anchor=north}}
\Tree    [.{ $\big\{ x_{0:5}^{(i)},  w^{(i)}_{0:5}  \big\}_{i = 1}^{N}  \sim p(x_{0:5}|y_{0:5}) =  \texttt{TS}(0,5)$} [. {$\big\{ x_{0:3}^{(i)},  w^{(i)}_{0:3}  \big\}_{i = 1}^{N} = \texttt{TS}{(0,3)}$}   [.$\big\{x_{0:1}^{(i)},w^{(i)}_{0:1}\big\}_{i = 1}^{N}=\texttt{TS}{(0,1)}$ [.$\big\{x_{0}^{(i)},w^{(i)}_{0}\big\}_{i = 1}^{N}$\\$=\texttt{TS}{(0, 0)}$ ] [.  $\big\{x_{1}^{(i)},w^{(i)}_{1}\big\}_{i = 1}^{N}$\\=$\texttt{TS}{(1 ,1)}$  ]] [.  $\big\{x_{2:3}^{(i)},w^{(i)}_{2:3}\big\}_{i = 1}^{N}=\texttt{TS}{(2, 3)}$   [.  $\big\{x_{2}^{(i)},w^{(i)}_{2}\big\}_{i = 1}^{N}$\\=$\texttt{TS}{(2, 2)}$  ] [. $\big\{x_{3}^{(i)},w^{(i)}_{3}\big\}_{i = 1}^{N}$\\$=\texttt{TS}{(3, 3)}$ ]]   ]   [. $\big\{x_{4:5}^{(i)},w^{(i)}_{4:5}\big\}_{i = 1}^{N}=\texttt{TS}{(4, 5)}$  [.$\big\{x_{4}^{(i)},w^{(i)}_{4}\big\}_{i = 1}^{N}$\\=$\texttt{TS}{(4, 4)}$ ]   [.    $\big\{x_{5}^{(i)},w^{(i)}_{5}\big\}_{i = 1}^{N}$\\=$\texttt{TS}{(5, 5)}$  ] ] ]   
\draw[gray, very thick,->] (7,-4.5) -- (7,0);
\end{tikzpicture}
\caption{Computational flow of $\texttt{TS}$ (see Algorithm \ref{RecursiveAlg}) in a HMM for $T = 5$. Each non-root node contains the weighted samples from the intermediate target distributions. The generation of the samples starts from the leaves following the branches towards the root of the auxiliary binary tree.}
\label{CF}
\end{figure}

Algorithm \ref{RecursiveAlg} demonstrates the generation of $N$ weighted samples $\big\{ x_{j:l}^{(i)},  w^{(i)}_{j:l}  \big\}$ from the target $f_{j:l}$ at $\mathcal{T}_{j:l}$.  
It adopts the pre-stored weighted particles $\big\{ \tilde{x}_{j:k-1}^{(i)},  \tilde{w}^{(i)}_{j:k-1}  \big\}_{i = 1}^{N}$ from $\mathcal{T}_{j:k-1}$ and $\big\{\tilde{x}_{k:l}^{(i)},  \tilde{w}^{(i)}_{k:l}  \big\}_{i = 1}^{N}$ from $\mathcal{T}_{k:l}$ where $k$ is the cutting point defined in Equation \eqref{cut_point}.  The algorithm first merges the weighted particles $\big \{\tilde{x}_{j:l}^{(i)}  = \big( \tilde{x}_{j:k-1}^{({i})}, \tilde{x}_{k:l}^{({i})} \big) \big\}_{i = 1}^{N}$ from the children which forms an approximation of the distribution with density $f_{j:k-1} f_{k:l}$.
The algorithm reweighs the combined samples using importance sampling to target the new distribution $f_{j:l}$.  
We retain the notation of the weights in the algorithm since some return unequal weights including Chopthin algorithm \citep{gandy2016chopthin} while others including multinomial resampling, residual resampling \citep{liu1998sequential} and systematic resampling \citep{kitagawa1996monte} return equal weights. 
We apply the algorithm recursively from the leaf nodes to the root of the auxiliary binary tree which yields the samples from the final target $f_{0:T} = p(x_{0:T} | y_{0:T})$. The computational flow is shown in Figure \ref{CF} when $T = 5$. 

The setting of the algorithms are the same as the paper by \cite{lindsten2017divide} with additional attentions to the form of the proposals and intermediate target distributions associated to the tree nodes. According to Proposition 1 and 2 in \cite{lindsten2017divide}, the unbiasedness of the normalising constant and the consistency can be verified under some regularity conditions given valid proposals and an exchangeable resampling procedure.  


\section{Intermediate target distributions in TPS} 
\label{target_distr}
Given an auxiliary tree $\mathcal{T}$ constructed in a way described in Section~\ref{Aux_tree}, we define the intermediate target distributions of the sub-models associated to the nodes in the tree.
We apply \cite{lindsten2017divide}'s method to build one class of intermediate target distribution $\{f_{j:l}\}_{\mathcal{T}_{j:l} \in \mathcal{T}}$ and develop two new classes, based on the
filtering and the smoothing distribution, respectively.

\subsection{Target suggested by \cite{lindsten2017divide}}
\cite{lindsten2017divide} recommends a class of intermediate target distributions with densities proportional to the product of the factors within the probabilistic graphical model. We apply the method to a HMM which bears binary and unary factors. A binary factor refers to a transition density of two consecutive hidden states. An unary factor refers to a prior density of a hidden state or the emission density between a hidden state and its observation. We call the tree-based particle smoothing algorithm with the above idea TPS-L as suggested by \cite{lindsten2017divide}.

At a leaf node $\mathcal{T}_{j}$ where the sub-model only contains a single random variable $X_{j}$ given the observation $Y_{j} = y_{j}$, the target distribution contains no binary factor and is defined as
 $f_{0}(x_{0}) \propto p_{0}(x_{0})p(y_{0} | x_{0})$ when $j = 0$ and $f_{j}(x_{j}) \propto p(y_{j} | x_{j})$ when $j \neq 0$.
 
At a non-leaf node $\mathcal{T}_{j:l}$, the target density is proportional to the product of all transition and emission densities containing the hidden states in the sub-model:
\begin{eqnarray*}
f_{j:l}(x_{j:l})  &\propto& p(y_{j} | x_{j})  \prod^{l-1}_{i = j} \bigg\{  p(x_{i+1} | x_{i})  p(y_{i+1} | x_{i+1})  \bigg\}.
\end{eqnarray*}
When $j = 0$, the prior density of $X_{0}$ is additionally multiplied.  

Assume $\mathcal{T}_{j:l}$ connects two children $\mathcal{T}_{j:k-1}, \mathcal{T}_{k:l} \in \mathcal{T}$ carrying the pre-generated particles: $\{ \tilde{x}^{(i)}_{j:k-1}, \tilde{w}^{(i)}_{j:k-1} \}_{i = 1}^{N} \sim f_{j:k-1}$ at $\mathcal{T}_{j:k-1} \in \mathcal{T}$ 
and $\{ \tilde{x}^{(i)}_{k:l}, \tilde{w}^{(i)}_{k:l} \}_{i = 1}^{N} \sim f_{k:l}$ at $\mathcal{T}_{k:l} \in \mathcal{T}$. 
The unnormalised importance weight $\hat{w}^{(i)}_{j:l}$ of the combined particle $\tilde{x}^{(i)}_{j:l} = (\tilde{x}_{j:k-1}^{({i})}, \tilde{x}_{k:l}^{({i})})$ in Equation \eqref{weight_formula} becomes:
\begin{eqnarray}
\hat{w}^{(i)}_{j:l} =  \tilde{w}_{j:l}^{(i)} p(\tilde{x}^{(i)}_{k} | \tilde{x}^{(i)}_{k-1}),
\end{eqnarray} 
where $\tilde{x}^{(i)}_{k-1}$ is the last element in  $\tilde{x}_{j:k-1}^{({i})}$ and $\tilde{x}^{(i)}_{k}$ is the first element in $\tilde{x}_{k:l}^{({i})}$. 

The tree-based sampling algorithm employing this type of intermediate target distributions is simple to implement, which does not involve any estimation techniques in the algorithms discussed in Section~\ref{inter_EF} and \ref{inter_ES}. TPS-L only requires the initial sampling of the particles from $f_{j}$ and applies importance sampling with a straightforward weight formula to merge them towards the root of the tree. The initial sampling distribution $f_{j}$ for $j \neq 0$ is equivalent to the posterior given a single observation $y_{j}$ from an uninformative prior. Correspondingly, the target distribution $\mathcal{T}_{j:l}$ only incorporates the observations from time $j$ to $l$ with no information beforehand or afterward.  We will see in the simulation section that with only one observation conditioned on, the initial sampling distribution may be vastly different from the marginal smoothing distribution, thus resulting in poor estimation results.

\subsection{Estimates of filtering distributions as target}
\label{inter_EF}
The second class of target distributions is based on estimates of
filtering distributions and thus we name the algorithm TPS-EF. At the
root,  the target distribution is
\begin{eqnarray*}
f_{0:T}(x_{0:T}) = p(x_{0:T}|y_{0:T}) = p_{0}(x_{0}) p(y_{0}|x_{0}) \prod^{T-1}_{i = 0} \bigg\{  p(x_{i+1} | x_{i})  p(y_{i+1} | x_{i+1})  \bigg\}.
\end{eqnarray*}
At a leaf node $\mathcal{T}_{j} \in \mathcal{T}$, we use an estimate of the filtering distribution $f_{j}(x_{j}) = \hat{p}(x_{j} | y_{0:j}) \approx p(x_{j} | y_{0:j})$ whose exact form and sampling process will be discussed in Section \ref{ini_target}. 
At a non-leaf and non-root node $\mathcal{T}_{j:l} \in \mathcal{T}$, we define the intermediate target distribution:
\begin{eqnarray*}
f_{j:l}(x_{j:l})  &\propto&  \hat{p}(x_{j} | y_{0:j})  \prod^{l-1}_{i = j} \bigg\{  p(x_{i+1} | x_{i})  p(y_{i+1} | x_{i+1})  \bigg\} \approx p(x_{j:l} | y_{0:l}).
\end{eqnarray*}
The weight of the merged sample $\tilde{x}^{(i)}_{j:l} = (\tilde{x}_{j:k-1}^{({i})}, \tilde{x}_{k:l}^{({i})})$ in Equation \eqref{weight_formula} becomes:
\begin{eqnarray}
\label{weight_1}
\hat{w}^{(i)}_{j:l} = \tilde{w}^{(i)}_{j:l} \frac{ p(  \tilde{x}_{k}^{({i})} |  \tilde{x}_{k-1}^{({i})}  )   p(y_{k} |   \tilde{x}_{k}^{({i})}) }{ \hat{p}_{k} (\tilde{x}^{({i})}_{k} | y_{0:k}) }.  
\end{eqnarray}


Under such constructions of the intermediate target distributions, the
particles at the leaf nodes are initially generated from (an estimate
of) the filtering distribution. Whilst moving up the tree, their
empirical marginal distributions gradually shifts towards the
smoothing distributions. One downside of this is that this may
eliminate a large population of particles, as the transition is
accomplished via importance sampling, particularly if the discrepancy
between the filtering and smoothing distribution is large.

\subsection{Kullback--Leibler divergence between the target and proposal distribution} 
\label{KLdiv}
Before proposing the second type of intermediate target distributions, we present an optimal type of proposal attaining the minimum Kullback--Leibler (KL) divergence \citep{cover2012elements} by assuming the random variables $X_{j:k-1} \in \mathcal{T}$ and $X_{k:l} \in \mathcal{T}$ from the sibling nodes being independent.  

Given the proposal $h_{j:l} = f_{j:k-1} f_{k:l}$ being the product of the densities of two independent random variables, the minimum KL divergence is met when the two densities are the marginals of the target densities with respect to the corresponding random variables. 
For simplicity of the notations, we denote the target density at a non-leaf node to be $f(\mathbf{x_{1}}, \mathbf{x_{2}})$ where $\mathbf{X_{1}}, \mathbf{X_{2}}$ are the random variables 
with the same time indices from the children but not necessarily the same probability measure. A valid proposal density $h_{1}( {\mathbf{x_{1}}})  h_{2}({ \mathbf{x_{2}}})$ satisfies $h_{1}( {\mathbf{x_{1}}})  h_{2}({ \mathbf{x_{2}}}) > 0$ whenever $f(\mathbf{x_{1}}, \mathbf{x_{2}}) > 0$, where we assume $h_{1}$ and $h_{2}$ are the probability densities of two independent (joint) random variables $\mathbf{X_{1}}$ and $\mathbf{X_{2}}$. We claim that proposal $f_{1}({\mathbf{x_{1}}})  f_{2}({ \mathbf{x_{2}}})$ has the smallest KL divergence among all proposals of the form $h_{1}(\mathbf{x_{1}})h_{2}({\mathbf{x_{2}}})$
where $f_{1}({\mathbf{x_{1}}})$ and $f_{2}({ \mathbf{x_{2}}})$ are the marginal densities of $f(\mathbf{x_{1}, x_{2}})$ with respect to $\mathbf{X_{1}}$ and $\mathbf{X_{2}}$, respectively.

\begin{theorem} 
\label{thm:KL_dst}

Let $f$ be a probability density function defined on $\mathbb{R}^{n_{1} + n_{2}}$, let $h_{1}$ and $h_{2}$ be probability density functions on $\mathbb{R}^{n_{1}}$ and $\mathbb{R}^{n_{2}}$, respectively. If $h_{1}(\mathbf{x_{1}}) h_{2} (\mathbf{x_{2}}) > 0$ whenever $f(\mathbf{x_{1}, x_{2}}) >0,$ then

$$ \int_{\mathbb{R}^{n_{2}}} \int_{\mathbb{R}^{n_{1}}}  f( \mathbf{x_{1}}, \mathbf{x_{2}}) \log \bigg(\frac{f(\mathbf{x_{1}}, \mathbf{x_{2}})}{h_{1}(\mathbf{x_{1}}) h_{2}(\mathbf{x_{2}}) }  \bigg) \mathrm{d} \mathbf{x_{1}} \mathrm{d}  \mathbf{x_{2}}  \geq 
\int_{\mathbb{R}^{n_{2}}} \int_{\mathbb{R}^{n_{1}}} f(\mathbf{x_{1}}, \mathbf{x_{2}}) \log \bigg(\frac{f(\mathbf{x_{1}}, \mathbf{x_{2}})}{f_{1}(\mathbf{x_{1}}) f_{2}(\mathbf{x_{2}}) }  \bigg) \mathrm{d} \mathbf{x_{1}} \mathrm{d}  \mathbf{x_{2}},$$
where
$f_{1}({ \mathbf{x_{1}}}) = \int_{\mathbb{R}^{n_{2}}}  f({\mathbf{x_{1}}}, \mathbf{{x_{2}}})  \mathrm{d}  { \mathbf{x_{2}}}$ and
$f_{2}({ \mathbf{x_{2}}}) =  \int_{\mathbb{R}^{n_{1}}}  f(\mathbf{{x_{1}}}, \mathbf{{x_{2}}})  \mathrm{d}  {\mathbf{x_{1}}}$
are the densities of the marginal distributions of $f(\mathbf{x_{1}}, \mathbf{x_{2}})$.  

\end{theorem}
The proof of Theorem \ref{thm:KL_dst} is in the Appendix.

\subsection{Estimates of smoothing distributions as target}
\label{inter_ES}
We provide an alternative way of constructing the intermediate target distributions using the marginal smoothing distributions motivated by Theorem \ref{thm:KL_dst}. Since the closed-form solutions to the marginal smoothing distributions are not available in general, we employ the estimates of the distributions at the nodes. At the root, we still use $f_{0:T} = p(x_{0:T}|y_{0:T})$. At a leaf node $\mathcal{T}_{j} \in \mathcal{T}$, we define $f_{j}(x_{j}) =  \hat{p}(x_{j} | y_{0:T}) \approx p(x_{j} | y_{0:T})$, which requires estimating the marginal smoothing distribution. We thus name the algorithm TPS-ES. 
At a non-leaf and non-root node $\mathcal{T}_{j:l}$, we define the target distribution $f_{j:l}$: 

\begin{eqnarray*}
f_{j:l}(x_{j:l})  &\propto&  \hat{p}(x_{j} | y_{0:j}) \frac{ \hat{p}(x_{l} | y_{0:T})}{ \hat{p}(x_{l} | y_{0:l})}  \prod^{l-1}_{i = j} \bigg\{  p(x_{i+1} | x_{i})  p(y_{i+1} | x_{i+1})  \bigg\} 
\\ &\approx&   p(x_{j} | y_{0:j})  \frac{p(x_{l} | y_{0:T}) }{p(x_{l} | y_{0:l}) }   \prod^{l-1}_{i = j} \bigg\{  p(x_{i+1} | x_{i})  p(y_{i+1} | x_{i+1})  \bigg\}\\
& = & p(x_{j:l} | y_{0:T}),
\end{eqnarray*}
where $\hat{p}(x_{j} | y_{0:j})$ denotes a probability density approximating the filtering density at the $j$th time step. Hence, given the estimate smoothing densities $\{ \hat{p}(x_{j} | y_{0:T}) \}_{j = 0, \ldots, T}$ and the estimating filtering densities $\{ \hat{p}(x_{j} | y_{0:j}) \}_{j = 0, \ldots, T}$, we build an estimator of the distribution $p(x_{j:l} | y_{0:T})$ at $\mathcal{T}_{j:l} \in \mathcal{T}$. Merging the particles at $\mathcal{T}_{j:l}$ from its children at $\mathcal{T}_{j:k-1} \in \mathcal{T}$ and $\mathcal{T}_{k:l} \in \mathcal{T}$ amounts to correlating the two sets of samples while roughly preserving  their marginal distributions. The weight of the merged sample $\tilde{x}^{(i)}_{j:l} = (\tilde{x}_{j:k-1}^{({i})}, \tilde{x}_{k:l}^{({i})})$ in Equation \eqref{weight_formula} becomes:

\begin{eqnarray} 
\label{weight_2}
\hat{w}^{(i)}_{j:l} =  \tilde{w}^{(i)}_{j:l} \frac{\hat{p}(\tilde{x}^{({i})}_{k-1}|y_{0:k-1})}{  \hat{p}(\tilde{x}^{({i})}_{k-1}|y_{0:T})  \hat{p}(\tilde{x}^{({i})}_{k}|y_{0:k})}  p(\tilde{x}^{({i})}_{k} | \tilde{x}^{({i})}_{k-1}) p(y_{k}|\tilde{x}^{({i})}_{k}).
\end{eqnarray}

Applying TPS-ES demands the constructions of $\{ \hat{p}(x_{j} | y_{0:j}) \}_{j = 0, \ldots, T}$ and $\{ \hat{p}(x_{j} | y_{0:T})  \}_{j = 0, \ldots, T}$ in advance. The new weight formula in Equation \eqref{weight_2} additionally incorporates the ratio between the estimated filtering and smoothing densities of $x_{k-1}$ compared with Equation \eqref{weight_1}. 

TPS-ES exhibits a sound property regarding the Kullback--Leibler divergence discussed in Section \ref{KLdiv}. Given the target distribution $f_{j:l}(x_{j:l}) = \hat{p}(x_{j:l}|y_{0:T}) $  estimating $p(x_{j:l}|y_{0:T}) $ at $\mathcal{T}_{j:l}$,  the proposal $h_{j:l}(x_{j:l}) = f_{j:k-1}(x_{j:k-1}) f_{k:l}(x_{k:l})$ estimates $p(x_{j:k-1} | y_{0:T}) p(x_{k:l} | y_{0:T})$.  We notice $p(x_{j:k-1} | y_{0:T})$ and $p(x_{k:l} | y_{0:T})$ are the marginal distributions, and their product forms a proposal attaining the minimum KL divergence from $p(x_{j:l}|y_{0:T})$. Hence, what the proposal density $h_{j:l}(x_{j:l})$ estimates has a minimum KL divergence from the smoothing density that our target distribution $f_{j:l}(x_{j:l})$ estimates.

Moreover, TPS-ES can be practically useful in some extreme models whereas the empirical marginal densities from other Monte Carlo smoothing algorithms may miss modes caused by the poor proposals. Since TPS-ES leaves the marginal distributions of all random variables roughly invariant at all levels of the tree, we can  diagnose each importance sampling step by inspecting the empirical marginals of the corresponding variables.  If there is a substantial difference between the empirical marginal distributions,  we need to examine the combination step.

\subsection{Initial sampling distribution at leaf nodes}
\label{ini_target}
We illustrate the constructions of the univariate distributions $\{\hat{p}(x_{j} | y_{1:j})\}_{j = 0, \ldots, T}$ and $\{\hat{p}(x_{j} | y_{0:T})_{j = 0, \ldots,T}\}$ mentioned in Section \ref{inter_EF} and Section \ref{inter_ES}, which are used in the initial sampling distributions at the leaf nodes. In general, the solutions of the filtering and smoothing distribution of a HMM are analytically intractable and need to be estimated from Monte Carlo samples with some exceptions including linear Gaussian and discrete HMMs. 

We aim to generate a probability density $\hat{f}$ estimating a target density $f$ given the weighted samples $\{x_{i}, w_{i}\}_{i = 1}^{n}$  from $f$. In the context of $f$ being a filtering or smoothing distribution, we can obtain the weighted samples by running a filtering algorithm or a smoothing algorithm. 
We are not interested in the empirical distribution since it is discrete and generally does not cover the full support of the random variable of interest. 

We first consider some parametric approaches. We can fit the data with some common probability distributions including a normal distribution and Student's $t$-distribution. We can also accommodate a mixture model to fit multiple modes of the target densities. The parameters of the distributions can be estimated in various ways including moment matching, maximum likelihood method and EM algorithm.  

The parametric approaches are reasonably quick and simple. For instance, assuming a Gaussian distribution requires the evaluation of the mean and variance and can be easily obtained from the samples using moment matching. The generation and evaluation of densities of the new particles are straightforward and fast to implement. Nevertheless, the target distribution may not be well approximated under the parametric assumption. 

Alternatively, we can employ some non-parametric approaches for instance, a kernel density estimator (KDS). We need to select the type of kernels and bandwidth in advance. The complexity of generating $N$ new samples is $O\big(\log(n)N\big)$ and the evaluation of the densities is more computationally expensive with complexity $O(nN)$.

We propose another non-parametric approximation method using piecewise constant functions with a lower computational effort than a KDS. We first build a uniform grid consisting of the points $x_{1} < x_{2} < \ldots < x_{n}$ with densities $d_{1}, \ldots, d_{n}$ estimated by a KDS such that $x_{i+1} - x_{i} = \Delta > 0$ for $i = 1, \ldots, n$.  The resulting probability density function formed by these grid points using piecewise constant functions is: 
\begin{align}
\label{interpolate}
f(x) = \sum_{i = 1}^{n}  \mathbbm{1}_{x \in [x_{i} - \Delta/2, x_{i} + \Delta/2)} d_{i}.
\end{align}
The evaluation of the sample densities reduces significantly from $O(nN)$ to $O\big(N \big)$ compared to a KDS.

Such probability density functions using piecewise constant functions have several disadvantages though enjoy a fast computation of estimated densities. Firstly, the estimator is biased since the proposal density generally does not cover the full support of the target density. Moreover, in TPS-ES, if the estimated filtering and smoothing distributions are both generated using the piecewise constant functions with different samples, there is no guarantee their densities have the same support, which may cause zero or infinite weight in Equation \eqref{weight_2}. To avoid this, we consider the mixture probability distributions using the piecewise constant functions accommodating the samples from both the filtering and smoothing distributions. Assume at time step $j$, the first uniform grid consists of the points $x^{f}_{1} < x^{f}_{2} < \ldots < x^{f}_{n^{f}}$ such that $x^{f}_{i+1} - x^{f}_{i} = \Delta^{f}$ for $i = 1, \ldots, n^{f}$ with estimated filtering densities $d^{f}_{1}, \ldots, d^{f}_{n}$ from a KDS and assume the second uniform grid consists of the points $x^{s}_{1} < x^{s}_{2} < \ldots < x^{s}_{n^{s}}$ such that $x^{s}_{i+1} - x^{s}_{i} = \Delta^{s}$ for $i = 1, \ldots, n^{s}$ with estimated smoothing densities $d^{s}_{1}, \ldots, d^{s}_{n}$ from another KDS. Then the resulting estimated filtering density $\hat{p}(x | y_{0:j})$ is given by 

\begin{align}
\label{filter_combine_smoother}
\hat{p}(x | y_{0:j}) = \alpha^{f} \sum_{i = 1}^{n^{f}}  \mathbbm{1}_{x \in [x^{f}_{i} - \Delta^{f}/2, x^{f}_{i} + \Delta^{f}/2)} d^{f}_{i} + (1-\alpha^{f}) \sum_{i = 1}^{n^{s}}  \mathbbm{1}_{x \in [x^{s}_{i} - \Delta^{s}/2, x^{s}_{i} + \Delta^{s}/2)} d^{s}_{i},
\end{align}
where $0 < \alpha^{f} < 1$.
Similarly, the estimated smoothing density $\hat{p}(x | y_{0:T})$ is given by 
\begin{align}
\label{smoother_combine_filter}
\hat{p}(x | y_{0:T}) = \alpha^{s} \sum_{i = 1}^{n^{s}}  \mathbbm{1}_{x \in [x^{s}_{i} - \Delta^{s}/2, x^{s}_{i} + \Delta^{s}/2)} d^{s}_{i} + (1-\alpha^{s}) \sum_{i = 1}^{n^{f}}  \mathbbm{1}_{x \in [x^{f}_{i} - \Delta^{f}/2, x^{f}_{i} + \Delta^{f}/2)} d^{f}_{i},
\end{align}
where $0 < \alpha^{s} < 1$. We have no conclusion of the values of $\alpha^{f}$ and $\alpha^{s}$ so far and choose them with values close to 1.
The resulting grid with the set of points $\{x^{f}_{1}, x^{f}_{2}, \ldots, x^{f}_{n^{f}}, x^{s}_{1}, x^{s}_{2}, \ldots, x^{s}_{n^{s}}\}$ is generally not uniform, but we ensure the estimated filtering and smoothing densities have the same support, though still finite.

\section{Simulations} 
\label{simulation}
We conduct simulations in a linear Gaussian HMM and a non-linear
non-Gaussian HMM in this section.  We implement TPS-EF and other
smoothing algorithms with roughly the same computational effort. In
the second example, we further compare TPS-EF and TPS-ES.

\subsection{Gaussian Linear Model} 
\label{linear_simulation}
We consider a simple linear Gaussian HMM similar to \citet{doucet2000sequential}.
\begin{alignat*}{2}
X_{t} &= 0.8 X_{t-1} +  V_{t} ~~~&& t = 1, \ldots, T,\\
Y_{t} &= X_{t} + W_{t} ~~~&& t = 0, \ldots, T.
\end{alignat*}
where $T = 127$, where $X_0,V_1,\dots,V_T,W_0,\dots,W_T$ are independent with $X_{0} \sim \mathcal{N}(0,1)$, $V_{t}\sim N(0,1)$,  $W_{t}\sim N(0,1)$.

We implement the following smoothing algorithms. We run TPS using
normal distributions as the initial sampling distributions (TPS-N)
whose means and variances are estimated using moment matching from the
samples of a bootstrap particle filter. The choice of a normal distribution is
motivated by the fact that in this case the true smoothing
distribution is a normal distribution. We also implement the tree-based particle smoothing algorithm suggest by \cite{lindsten2017divide} (TPS-L), the
Rauch--Tung--Striebel smoother (RTSs) \citep{rauch1965maximum}
yielding the closed-form solutions, the bootstrap particle filter
(BPF) which updates the entire history of the particles in each step, the
forward filtering backward smoothing algorithm (FFBSm)
\citep{doucet2000sequential} and the forward filtering backward
simulation (FFBSi) \citep{godsill2004monte}.

We have implemented the above methods in \texttt{R} ourselves. We set the required sample size $N = 10000$ in TPS-N as a benchmark and denote $n = 10000$ the number of samples pre-generated from a bootstrap particle filter in FFBSm, FFBSi, TPS-N and TPS-L. We adjust the number of particles in other algorithms to roughly keep the same running time. As the implementations are not deterministic, we allow a 10\% error regarding the running time for the rest of the algorithms compared to TPS-N. We run each algorithm $M = 500$ times with the same set of observations $\{y_{t}\}_{t = 0}^{127}$.  

As a criterion for comparison, we define the mean square error of means (MSEm) and variances (MSEv) in the $m$th simulation:

\begin{eqnarray*}
\text{MSEm}_{m} &=&   \frac{1}{T+1}  \sum^{T}_{t = 0} \big(  \widehat{\mathbb{E}}^{m}[X_{t}| Y_{0:T}] - \mathbb{E}[X_{t}| Y_{0:T}]\big)^{2},  \\
\text{MSEv}_{m} &=&    \frac{1}{T+1}  \sum^{T}_{t = 0} \big(  \widehat{\text{Var}}^{m}[X_{t}| Y_{0:T}] - \text{Var}[X_{t}| Y_{0:T}]\big)^{2},   \\
\end{eqnarray*}
where $ \widehat{\mathbb{E}}[X^{m}_{t}| Y_{0:T}]$ and $\widehat{\text{Var}}[x^{m}| y_{0:T}] $ are the Monte Carlo estimates of the mean and variance of the smoothing distribution at time step $t$ in the $m$th simulation. $\mathbb{E}[X_{t}| Y_{0:T}]$ and $\text{Var}[X_{t}| Y_{0:T}]$ are the true smoothing means and variances from a Rauch--Tung--Striebel smoother \citep{rauch1965maximum}. 

The simulation results are shown in Table \ref{tablelinear}. When $N = n$, the two tree-based sampling algorithms: TPS-L and TPS-N enjoy the same complexity $O(N)$ as BPF, and generate far more particles than FFBSm and FFBSi with quadratic complexities. TPS-L has the smallest mean of MSEm and MSEv followed by TPS-N, which outperform FFBSm and FFBSi significantly in terms of MSEm.

\begin{table}[ht]
\centering
\caption{Simulation errors in the linear model}
\label{tablelinear}
\begin{tabular}{rrrrrrr}
  \hline
 & $N$ & $n$ & Mean of MSEm (s.e.) & Mean of MSEv (s.e.) \\ 
  \hline
  BPF & 44000 & NA & 0.0020 (0.0000147) & 0.0019 (0.000013) \\ 
  FFBSm & 410 & 410 & 0.0065 (0.0000550) & 0.0047 (0.000037) \\ 
  FFBSi & 450 & 450 & 0.0059 (0.0000563) & 0.0044 (0.000031) \\ 
  TPS-N & 10000 & 10000 & 0.0014 (0.0000096) & 0.0018 (0.000014) \\ 
  TPS-L & 13000 & NA & 0.0008 (0.0000061) & 0.0007 (0.000005) \\ 
 \hline
\end{tabular}
\end{table}

\subsection{Non-linear Model} 
\label{non-linear_simulation}
We consider a well-known non-linear model \citep{gordon1993novel, andrieu2010particle}:

\begin{alignat*}{2}
X_{t} &= \frac{1}{2} X_{t-1} + 25 \frac{X_{t-1}}{1 + X_{t-1}^2} + 8 \cos (1.2 t) + V_{t},~~&&t = 1,2, \ldots, T,\\
Y_{t} &= \frac{X^{2}_{t}}{20} + W_{t},~~&&t = 0,2, \ldots, T,
\end{alignat*}
where $T = 511$, where $X_0, V_1,...,V_T, W_0,...W_T$ are independent with $X_{0} \sim \mathcal{N}(0, 1)$,  $V_{t}  \sim \mathcal{N}(0, \tau^2)$ and $W_{t}  \sim \mathcal{N}(0,\sigma^2).$  
 
We run the same algorithms BPF, FFBSm, FFBSi and TPS-L as in Section
\ref{linear_simulation}. In TPS-EF, we use piecewise constant functions
defined in Equation \eqref{interpolate} for the approximation of the
initial sampling distributions. We call the algorithm TPS-EFP and set
$N = n = 10000$ as a benchmark. As before, we correspondingly adjust
the sample sizes in other algorithms to achieve roughly the same
computational effort.

We calculate the mean and standard deviation of the MSE of means (MSEm) in $M = 500$ simulations with the same set of observations. Given no closed-form solutions to the true smoothing distributions, we apply a discrete analogue to the distributions of the initial hidden state $p_{0}(x_{0})$ and the transition distributions $\{p(x_{t+1}|x_{t})\}_{t = 0, \ldots, 126}$.
We then approximate the smoothing distributions of the original HMM using the solutions of the discrete-space HMM.  The MSEm of the $m$th simulation in the non-linear model is defined as:

$$  \text{MSEm}_{m}  =  \frac{1}{T+1}  \sum^{T}_{t = 0} \big(\widehat{\mathbb{E}}^{m}[X_{t}| Y_{0:T}] - \mathbb{E} (\hat{X}_{t}\big | y_{0:T}) )^{2}, $$
where $\mathbb{E} (\hat{X}_{t}\big | y_{0:T})$ is the mean of the smoothing distribution at time step $t$ of the discrete-space HMM.

We additionally perform Kolmogorov--Smirnov test \citep{massey1951kolmogorov} which measures a distance between the empirical distribution and the target probability distribution.
In the context of the smoothing problem in a non-linear hidden Markov model, the Kolmogorov--Smirnov statistic can be defined as

$$ D = \sup _{x} |  F^{(t)}_{1,N}(x) -  F^{(t)}_{2}(x) |, $$
where $F^{(t)}_{1,N}$ is the empirical cumulative function generated by $N$ samples at the time step $t$ from a smoothing algorithm and $F^{(t)}_{2}$ is the cumulative distribution function at time step $t$ of the smoothing distribution from a discrete-space HMM derived from the true model. We denote KS$_{m}$ to be the sum of the KS statistic of all time steps in the $m$th simulation.

\begin{figure}
\centering
\includegraphics[width=0.8\textwidth]{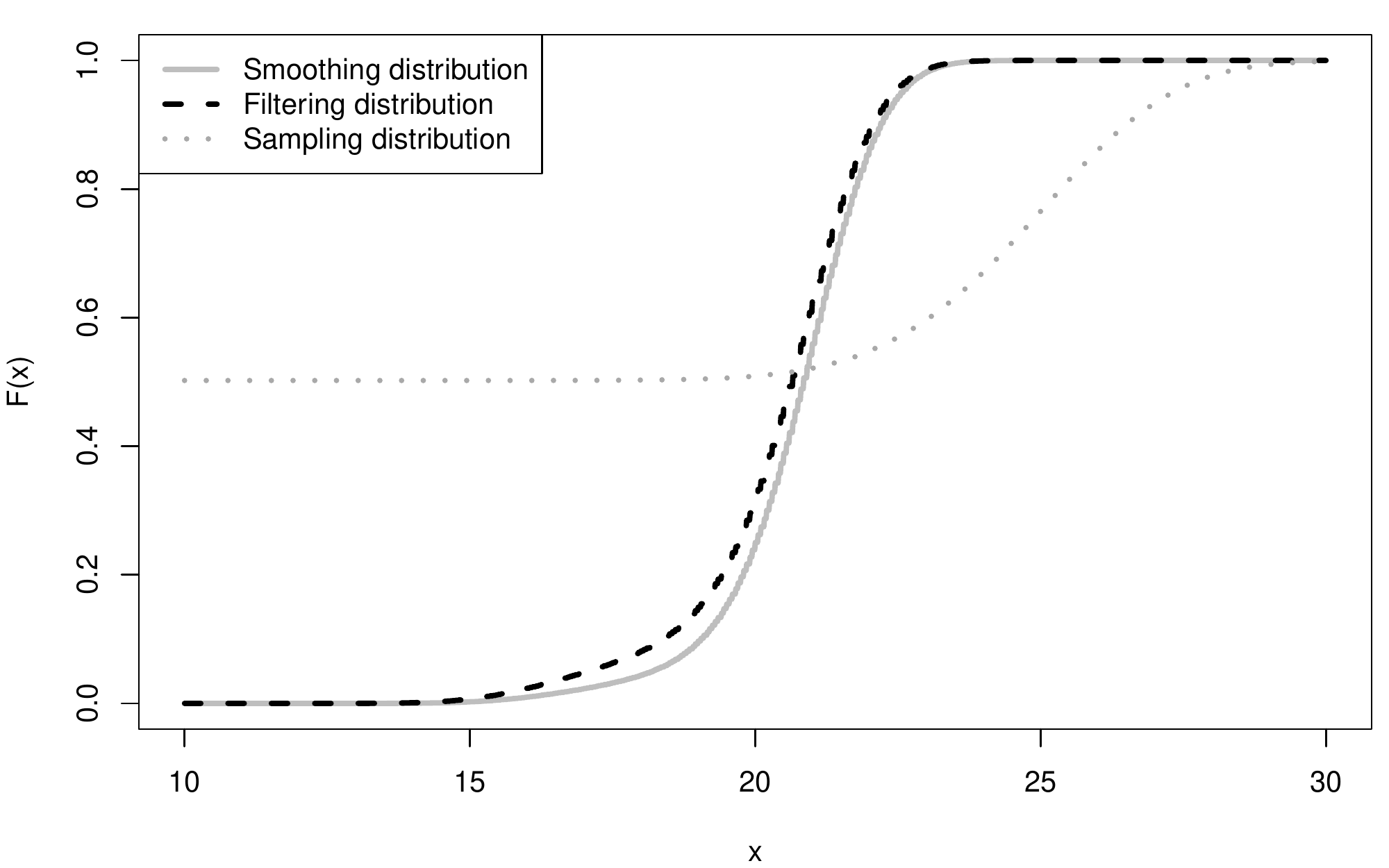}
\caption{CDF of the smoothing, filtering and initial sampling distribution at time step $j = 271$ of TPS-L in the non-linear model when $\tau = 1, \sigma = 5$.}
\label{fig:ecdf272}
\end{figure}

\begin{table}[tbp]
\centering
\caption{Simulation errors in the non-linear model}
\label{tablenon-linear}
\begin{tabular}{rrrrrrrr}
  \hline
 & Parameter Values & $N$ & $n$ &  Mean of MSEm (s.e.) & Mean of KS \\ 
  \hline
\multicolumn{1}{c|}{BPF}  & \multicolumn{1}{c|}{\multirow{4}{*}{$\tau = 1, \sigma = 1$}} & 40000 & NA &  0.0239 (0.00085) & 80.04 \\ 
\multicolumn{1}{c|}{FFBSm}  & \multicolumn{1}{c|}{} & 315 & 315 &  0.0944 (0.01657) & 77.20 \\ 
\multicolumn{1}{c|}{FFBSi}  & \multicolumn{1}{c|}{} & 320 & 320 &  0.1399 (0.02291) & 76.65 \\ 
\multicolumn{1}{c|}{TPS-EFP}  & \multicolumn{1}{c|}{} & 10000 & 10000 &  0.0050 (0.00007) & 34.51 \\ 
\multicolumn{1}{c|}{TPS-L}  & \multicolumn{1}{c|}{} & 13000 & NA &  0.3020 (0.00042) & 109.13 \\
\cline{2-2}
\multicolumn{1}{c|}{BPF}  & \multicolumn{1}{c|}{\multirow{4}{*}{$\tau = 1, \sigma = 5$}} & 40000 & NA &  0.2096 (0.03064) & 55.10 \\ 
\multicolumn{1}{c|}{FFBSm}  & \multicolumn{1}{c|}{} & 315 & 315 &  0.6785 (0.02850) & 67.71 \\ 
\multicolumn{1}{c|}{FFBSi}  & \multicolumn{1}{c|}{} & 320 & 320 &  0.6071 (0.04981) & 66.12 \\ 
\multicolumn{1}{c|}{TPS-EFP}  & \multicolumn{1}{c|}{} & 10000 & 10000 &  0.3998 (0.01174) & 47.36 \\
\multicolumn{1}{c|}{TPS-L}  & \multicolumn{1}{c|}{} & 13000 & NA &  14.4847 (0.01790) & 261.34 \\
\cline{2-2}
\multicolumn{1}{c|}{BPF}  & \multicolumn{1}{c|}{\multirow{4}{*}{$\tau = 5, \sigma = 1$}} & 40000 & NA & 1.2182 (0.05684) & 119.33 \\ 
\multicolumn{1}{c|}{FFBSm}  & \multicolumn{1}{c|}{} & 315 & 315 &  3.4342 (0.22357) & 94.57 \\ 
\multicolumn{1}{c|}{FFBSi}  & \multicolumn{1}{c|}{} & 320 & 320 &  3.2161 (0.20196) & 93.60 \\ 
\multicolumn{1}{c|}{TPS-EFP}  & \multicolumn{1}{c|}{} & 10000 & 10000 &  0.1034 (0.00544) & 28.19 \\ 
\multicolumn{1}{c|}{TPS-L}  & \multicolumn{1}{c|}{} & 13000 & NA &  0.4599 (0.00149) & 67.69 \\\cline{2-2}
   \hline
\end{tabular}
\end{table}
The simulation results with different values of $\tau$ and $\sigma$ are shown in Table \ref{tablenon-linear}. In the first two situations, TPS-L shows the largest error and KS statistic, especially when $\tau = 1$ and $\sigma = 5$. 
This can be explained by the poor proposal from the initial sampling distribution constructed by the algorithm.  We examine this by plotting the cumulative distribution function (CDF) of the initial sampling distribution $f_{j}$ in TPS-L, the filtering distribution $p(x_{j}| y_{0:j})$ and the marginal smoothing distribution $p(x_{j}| y_{0:T})$ at a particular time step when $j = 271$.  In Figure \ref{fig:ecdf272}, the CDF of the initial sampling are far more dissimilar to the marginal smoothing distribution than the filtering one, which contributes to very ineffective importance sampling steps during the built-up of the tree.

Other algorithms provide different results in the three parameter settings. When $\tau = 1, \sigma = 1$, TPS-EFP shows much smaller MSEm and KS followed by BPF. BPF however has the largest mean of KS. 
When $\tau = 1, \sigma = 5$, TPS-EFP has a larger mean of MSEm than BPF. In terms of the KS statistic, TPS-EFP outperforms other smoothing algorithms. When $\tau = 5, \sigma = 1$, TPS-EFP and TPS-L produce dominant results with vastly smaller MSEm. They also exhibit the smallest mean of KS among the smoothing algorithms whereas the BPF gives the largest result though generating the most samples. 

To conclude, TPS-EFP and TPS-L perform well when the ratio between the standard deviation in the transition and emission density, i.e. when $\tau / \sigma$  is large.
TPS-EFP has a more stable and appreciable performance, which provides low MSEm and consistently the smallest KS among the five smoothing algorithms. In contrast, the result of TPS-L may be misleading due to its instability. BPF works well regarding MSEm in some situations, but poorly in terms of Kolmogorov--Smirnov statistic. FFBSm and FFBSi produces less accurate results due to higher computational complexity.

\subsection{Comparing TPS-EF and TPS-ES in the non-linear model}
In this section, we conduct simulations in the same non-linear model using tree-based particle smoothing algorithm with estimated filtering (TPS-EF) and smoothing (TPS-ES) distributions as the intermediate target distributions. As TPS-ES is not a good competitor given a relatively small sample size in Section \ref{non-linear_simulation}, we compare its performance with TPS-EF with more computational budget.
 
We demonstrate the implementations of the two algorithms. We apply the same smoothing algorithm TPS-EFP as described in Section \ref{non-linear_simulation} which utilises the piecewise constant functions to estimate the filtering distributions. As TPS-ES requires the estimated smoothing distributions as the initial sampling distributions, we achieve this by using piecewise constant functions for the estimation based on the samples from an initial run of TPS-EFP and thus call the algorithm TPS-ESP. 

We specify the parameters in the simulations of TPS-EFP and TPS-ESP. We denote the sample size by $N$. We set the parameters $\alpha^{s} = \alpha^{f} = 0.95$ appeared in Equation \eqref{filter_combine_smoother} and \eqref{smoother_combine_filter}. In TPS-EFP and TPS-ESP, the estimated filtering distributions are both constructed from $n$ samples from the particle filters. Additionally, in TPS-ESP, the estimated smoothing distributions are constructed from TPS-EFP with $n'$ samples. We run TPS-ESP in two different situations: The first one has the same sample size $N$ as TPS-EFP and requires more computational effort to estimate the initial sampling distributions based on $n'$ Monte Carlo samples. The second one has roughly the same computational effort as TPS-EFP which generates fewer Monte Carlo samples for the estimation of the initial sampling distributions and the target samples.

We compare TPS-EFP and TPS-ESP with respect to the mean square error and Kolmogorov--Smirnov statistic defined in Section \ref{non-linear_simulation}.  We run TPS-EFP and TPS-ESP for $M = 200$ times with different values of $\tau$ and $\sigma$ whose results are shown in Table \ref{TPS_compare}. 
TPS-ESP has an evident improvement of the Kolmogorov--Smirnov (KS) statistic in most situations and the comparisons between MSEm vary. The MSEm of TPS-ESP always decreases when generating the same number of samples as TPS-EFP. However, TPS-ESP does not provide convinced results under roughly the same computational effort.

Overall, the performance of TPS-ESP depends on the computational budget. Given the same sample size as in TPS-EFP, TPS-ESP can potentially decrease both MSEm and KS statistic. This may not be true when the algorithm is kept the same overall effort as TPS-EFP.

\begin{table}[tbp]
\centering
\caption{Simulation errors between TPS-EF and TPS-ES in the non-linear model}
\label{TPS_compare}
\begin{tabular}{rrrrrrrr}
  \hline
 & Parameter Values & $N$ & $n$ & $n'$ & Mean of MSEm (s.e.) & Mean of KS \\ 
  \hline
\multicolumn{1}{c|}{TPS-EFP}  & \multicolumn{1}{c|}{\multirow{3}{*}{$\tau = 1,\sigma = 1$}} & 50000 & 50000 & NA & 0.00123 (0.00032) & 17.56 \\ 
\multicolumn{1}{c|}{TPS-ESP}  & \multicolumn{1}{c|}{} & 50000 & 50000 & 50000 & 0.00051 (0.00007) & 11.91 \\ 
\multicolumn{1}{c|}{TPS-ESP}  & \multicolumn{1}{c|}{} & 18000 & 50000 & 25000 & 0.00169 (0.00037) & 15.17 \\  \cline{2-2}
\multicolumn{1}{c|}{TPS-EFP}  & \multicolumn{1}{c|}{\multirow{3}{*}{$\tau = 1,\sigma = 5$}} & 50000 & 50000 & NA & 0.09136 (0.02758) & 24.27 \\ 
\multicolumn{1}{c|}{TPS-ESP}  & \multicolumn{1}{c|}{} & 50000 & 50000 & 50000 & 0.10297 (0.01128) & 19.51 \\ 
\multicolumn{1}{c|}{TPS-ESP}  & \multicolumn{1}{c|}{} & 18000 & 50000 & 25000 & 0.19861 (0.01954) & 26.56\\   \cline{2-2}
\multicolumn{1}{c|}{TPS-EFP}  & \multicolumn{1}{c|}{\multirow{3}{*}{$\tau = 5,\sigma = 1$}} & 50000 & 50000 & NA & 0.01420 (0.01193) & 14.63 \\ 
\multicolumn{1}{c|}{TPS-ESP}  & \multicolumn{1}{c|}{} & 50000 & 50000 & 50000 & 0.01261 (0.00269) & 11.87 \\ 
\multicolumn{1}{c|}{TPS-ESP}  & \multicolumn{1}{c|}{} & 18000 & 50000 & 25000 & 0.02599 (0.00509) & 14.80 \\  \cline{2-2}
   \hline
\end{tabular}
\end{table}



\section{Conclusion}
\label{conclusion}
This article introduces a Monte Carlo sampling method we call TPS built on the D\&C SMC \citep{lindsten2017divide} to estimate the joint smoothing distribution $p(x_{0:T}|y_{0:T})$ in a hidden Markov model. The method decomposes the model into sub-models with intermediate target distributions using a binary tree structure. TPS samples independently from the leaves of the tree and gradually merges and resamples to target the new distributions upon the auxiliary tree.

We propose one generic way of constructing a binary tree which sequentially splits the joint random variables $X_{0:T}$. Furthermore, we discuss the sampling procedure of the target samples at a non-leave node by combining the samples from its children using importance sampling. The computational effort is adjustable with a possible reduction to a linear effort with respect to the sample size. 

Using the above settings, we investigate three algorithms with different types of intermediate target distributions at the non-root nodes. TPS-L \citep{lindsten2017divide} constructs intermediate target distributions conditional on the observations from the same time interval as the target variables and imposes an uninformative prior. TPS-L is very simple to implement with no additional tuning algorithms. The algorithm is at the risk of providing very poor initial sampling distribution based on little information from the observations. TPS-EF employs intermediate target distributions estimating of the (joint) filtering distributions which conditions on the observations up to the last time step in the target variable. 
It is straightforward for implementation with an initial run of a filtering algorithm. Nevertheless, the proposal in the importance sampling step may still not be satisfactory when the marginal filtering and smoothing distributions are vastly different. TPS-ES builds the distributions estimating of the (joint) smoothing distributions which conditions on all the observations. It roughly retains the marginal smoothing distributions from the intermediate target distributions at all levels of the auxiliary tree despite its more intensive computations.

We further propose the constructions of the estimated filtering and smoothing distributions based on the Monte Carlo samples. Considering both accuracy and computational effort, we recommend parametric approaches such as normal assumptions in a linear Gaussian model and non-parametric approaches such as using piecewise constant functions in a non-linear model.  

In the simulation studies, TPS-L has the smallest error in the linear model, but very unstable results in the different settings of the non-linear model. TPS-EF exhibits more desirable simulation outcomes. It is computationally less expensive than the most smoothing algorithms with quadratic complexity. It also produces the smallest mean square errors in the linear Gaussian model and consistently the smallest average Kolmogorov--Smirnov statistic in different situations under the non-linear model. In particular, it outperforms other algorithms substantially when the variance of the transition density is much larger than the emission density. TPS-ES, however, has a better approximation of the smoothing distribution with respect to the Kolmogorov--Smirnov statistic compared with TPS-EF at the cost of an additional run of a smoothing algorithm.

To conclude, TPS with two proposed choices of the intermediate target distribution presents a new approach of addressing the smoothing problem which shows the following advantages: We have flexibilities of choosing and constructing the intermediate target distributions, which can potentially produce better proposals in the importance sampling steps. TPS can escape from the quadratic complexity with respect to the sample size computationally, and produce more particles and accurate simulation results than some smoothing algorithms. Nevertheless, its performance depends on the implementation of other filtering or smoothing algorithms and the estimation of the target distributions. Due to its flexible and relatively fast implementations with stable and comparable simulation results, we regard it as a competitor with other smoothing algorithms.





\bibliographystyle{chicago}

\begin{thebibliography}{}

\bibitem[\protect\citeauthoryear{Andrieu, Doucet, and Holenstein}{Andrieu
  et~al.}{2010}]{andrieu2010particle}
Andrieu, C., A.~Doucet, and R.~Holenstein (2010).
\newblock Particle {M}arkov chain {M}onte {C}arlo methods.
\newblock {\em Journal of the Royal Statistical Society: Series B (Statistical
  Methodology)\/}~{\em 72\/}(3), 269--342.

\bibitem[\protect\citeauthoryear{Arulampalam, Maskell, Gordon, and
  Clapp}{Arulampalam et~al.}{2002}]{arulampalam2002tutorial}
Arulampalam, M.~S., S.~Maskell, N.~Gordon, and T.~Clapp (2002).
\newblock A tutorial on particle filters for online nonlinear/non-{G}aussian
  {B}ayesian tracking.
\newblock {\em Signal Processing, IEEE Transactions on\/}~{\em 50\/}(2),
  174--188.

\bibitem[\protect\citeauthoryear{Baum and Petrie}{Baum and
  Petrie}{1966}]{baum1966statistical}
Baum, L.~E. and T.~Petrie (1966).
\newblock Statistical inference for probabilistic functions of finite state
  {M}arkov chains.
\newblock {\em The Annals of Mathematical Statistics\/}~{\em 37\/}(6),
  1554--1563.

\bibitem[\protect\citeauthoryear{Beskos, Jasra, Law, Tempone, and Zhou}{Beskos
  et~al.}{2017}]{beskos2017multilevel}
Beskos, A., A.~Jasra, K.~Law, R.~Tempone, and Y.~Zhou (2017).
\newblock Multilevel sequential {M}onte {C}arlo samplers.
\newblock {\em Stochastic Processes and their Applications\/}~{\em 127\/}(5),
  1417--1440.

\bibitem[\protect\citeauthoryear{Briers, Doucet, and Maskell}{Briers
  et~al.}{2010}]{briers2010smoothing}
Briers, M., A.~Doucet, and S.~Maskell (2010).
\newblock Smoothing algorithms for state--space models.
\newblock {\em Annals of the Institute of Statistical Mathematics\/}~{\em
  62\/}(1), 61--89.

\bibitem[\protect\citeauthoryear{Cover and Thomas}{Cover and
  Thomas}{2012}]{cover2012elements}
Cover, T.~M. and J.~A. Thomas (2012).
\newblock {\em Elements of information theory}.
\newblock John Wiley \& Sons.

\bibitem[\protect\citeauthoryear{Doucet, De~Freitas, and Gordon}{Doucet
  et~al.}{2001}]{de2001sequential}
Doucet, A., N.~De~Freitas, and N.~Gordon (2001).
\newblock {\em Sequential {M}onte {C}arlo methods in practice}.
\newblock Springer.

\bibitem[\protect\citeauthoryear{Doucet, Godsill, and Andrieu}{Doucet
  et~al.}{2000}]{doucet2000sequential}
Doucet, A., S.~Godsill, and C.~Andrieu (2000).
\newblock On sequential {M}onte {C}arlo sampling methods for {B}ayesian
  filtering.
\newblock {\em Statistics and Computing\/}~{\em 10\/}(3), 197--208.

\bibitem[\protect\citeauthoryear{Fearnhead, Wyncoll, and Tawn}{Fearnhead
  et~al.}{2010}]{fearnhead2010sequential}
Fearnhead, P., D.~Wyncoll, and J.~Tawn (2010).
\newblock A sequential smoothing algorithm with linear computational cost.
\newblock {\em Biometrika\/}~{\em 97\/}(2), 447--464.

\bibitem[\protect\citeauthoryear{Gandy and Lau}{Gandy and
  Lau}{2016}]{gandy2016chopthin}
Gandy, A. and F.~D.-H. Lau (2016).
\newblock The chopthin algorithm for resampling.
\newblock {\em IEEE Trans. Signal Processing\/}~{\em 64\/}(16), 4273--4281.

\bibitem[\protect\citeauthoryear{Gerber and Chopin}{Gerber and
  Chopin}{2015}]{gerber2015sequential}
Gerber, M. and N.~Chopin (2015).
\newblock Sequential quasi {M}onte {C}arlo.
\newblock {\em Journal of the Royal Statistical Society: Series B (Statistical
  Methodology)\/}~{\em 77\/}(3), 509--579.

\bibitem[\protect\citeauthoryear{Godsill, Doucet, and West}{Godsill
  et~al.}{2004}]{godsill2004monte}
Godsill, S.~J., A.~Doucet, and M.~West (2004).
\newblock {M}onte {C}arlo smoothing for nonlinear time series.
\newblock {\em Journal of the {A}merican {S}tatistical {A}ssociation\/}~{\em
  99\/}(465).

\bibitem[\protect\citeauthoryear{Gordon, Salmond, and Smith}{Gordon
  et~al.}{1993}]{gordon1993novel}
Gordon, N.~J., D.~J. Salmond, and A.~F. Smith (1993).
\newblock Novel approach to nonlinear/non-gaussian bayesian state estimation.
\newblock {\em IEE Proceedings F (Radar and Signal Processing)\/}~{\em
  140\/}(2), 107--113.

\bibitem[\protect\citeauthoryear{Kitagawa}{Kitagawa}{1987}]{kitagawa1987non}
Kitagawa, G. (1987).
\newblock Non-{G}aussian state space modeling of nonstationary time series.
\newblock {\em Journal of the {A}merican {S}tatistical {A}ssociation\/}~{\em
  82\/}(400), 1032--1041.

\bibitem[\protect\citeauthoryear{Kitagawa}{Kitagawa}{1996}]{kitagawa1996monte}
Kitagawa, G. (1996).
\newblock {M}onte {C}arlo filter and smoother for non-{G}aussian nonlinear
  state space models.
\newblock {\em Journal of Computational and Graphical Statistics\/}~{\em
  5\/}(1), 1--25.

\bibitem[\protect\citeauthoryear{Klaas, Briers, De~Freitas, Doucet, Maskell,
  and Lang}{Klaas et~al.}{2006}]{klaas2006fast}
Klaas, M., M.~Briers, N.~De~Freitas, A.~Doucet, S.~Maskell, and D.~Lang (2006).
\newblock Fast particle smoothing: {I}f {I} had a million particles.
\newblock In {\em Proceedings of the 23rd international conference on Machine
  learning}, pp.\  481--488. ACM.

\bibitem[\protect\citeauthoryear{Lindsten, Johansen, Naesseth, Kirkpatrick,
  Sch{\"o}n, Aston, and Bouchard-C{\^o}t{\'e}}{Lindsten
  et~al.}{2017}]{lindsten2017divide}
Lindsten, F., A.~M. Johansen, C.~A. Naesseth, B.~Kirkpatrick, T.~B. Sch{\"o}n,
  J.~Aston, and A.~Bouchard-C{\^o}t{\'e} (2017).
\newblock Divide-and-conquer with sequential {M}onte {C}arlo.
\newblock {\em Journal of Computational and Graphical Statistics\/}~{\em
  26\/}(2), 445--458.

\bibitem[\protect\citeauthoryear{Liu and Chen}{Liu and
  Chen}{1998}]{liu1998sequential}
Liu, J.~S. and R.~Chen (1998).
\newblock Sequential {M}onte {C}arlo methods for dynamic systems.
\newblock {\em Journal of the American Statistical Association\/}~{\em
  93\/}(443), 1032--1044.

\bibitem[\protect\citeauthoryear{Massey~Jr}{Massey~Jr}{1951}]{massey1951kolmogorov}
Massey~Jr, F.~J. (1951).
\newblock The {K}olmogorov-{S}mirnov test for goodness of fit.
\newblock {\em Journal of the American statistical Association\/}~{\em
  46\/}(253), 68--78.

\bibitem[\protect\citeauthoryear{Naesseth, Linderman, Ranganath, and
  Blei}{Naesseth et~al.}{2017}]{naesseth2017variational}
Naesseth, C.~A., S.~W. Linderman, R.~Ranganath, and D.~M. Blei (2017).
\newblock Variational sequential {M}onte {C}arlo.
\newblock {\em arXiv preprint arXiv:1705.11140\/}.

\bibitem[\protect\citeauthoryear{Rauch, Striebel, and Tung}{Rauch
  et~al.}{1965}]{rauch1965maximum}
Rauch, H.~E., C.~Striebel, and F.~Tung (1965).
\newblock Maximum likelihood estimates of linear dynamic systems.
\newblock {\em AIAA Journal\/}~{\em 3\/}(8), 1445--1450.

\end{thebibliography}

\appendix
\section{Proof of Theorem \ref{thm:KL_dst}}

By Jensen's inequality, 
\begin{eqnarray*}
&& \int_{\mathbb{R}^{n_{1}}}  f_{1}(\mathbf{x_{1}}) \log \big( f_{1}(\mathbf{x_{1}}) \big)  \mathrm{d} \mathbf{x_{1}} -    \int_{\mathbb{R}^{n_{1}}}  f_{1}(\mathbf{x_{1}})  \log \big(  h_{1}(\mathbf{x_{1}})  \big) \mathrm{d}\mathbf{x_{1}}\\ &=&    \int_{\mathbb{R}^{n_{1}}}   f_{1} (\mathbf{x_{1}}) \log \bigg(   \frac{f_{1}(\mathbf{x_{1}})} {h_{1} (\mathbf{x_{1}})} \bigg)  \mathrm{d} \mathbf{x_{1}}
 =  \mathbb{E}   \bigg[ \log \bigg( \frac{f_{1}(\mathbf{X_{1}})}{ h_{1}(\mathbf{X_{1}}) }   \bigg)  \bigg] =  \mathbb{E}   \bigg[  - \log \bigg( \frac{h_{1}(\mathbf{X_{1}})}{ f_{1}(\mathbf{X_{1}}) }   \bigg)  \bigg] \\
&\geq& - \log \bigg\{   \mathbb{E}  \bigg[  \frac{h_{1}(\mathbf{X_{1}})}{f_{1}(\mathbf{X_{1}})} \bigg]   \bigg\} = 0. \end{eqnarray*} 

Using this and the definition of marginal distribution, 
\begin{eqnarray}
\label{eq:entropy1}
&&  \int_{\mathbb{R}^{n_{2}}}  \int_{\mathbb{R}^{n_1}}    f(\mathbf{x_{1}}, \mathbf{x_{2}}) \log \big( f_{1}(\mathbf{x_{1}}) \big)  \mathrm{d} \mathbf{x_{1}} \mathrm{d} \mathbf{x_{2}} 
=  \int_{\mathbb{R}^{n_{1}}}  f_{1}(\mathbf{x_{1}}) \log \big( f_{1}(\mathbf{x_{1}}) \big)  \mathrm{d} \mathbf{x_{1}} \nonumber \\ 
&\geq& \int_{\mathbb{R}^{n_{1}}}  f_{1}(\mathbf{x_{1}}) \log \big( h_{1}(\mathbf{x_{1}}) \big)  \mathrm{d} \mathbf{x_{1}} = \int_{\mathbb{R}^{n_{2}}} \int_{\mathbb{R}^{n_1}}    f(\mathbf{x_{1}}, \mathbf{x_{2}})  \log \big(  h_{1}(\mathbf{x_{1}})  \big) \mathrm{d} \mathbf{x_{1}}\mathrm{d}  \mathbf{x_{2}}.  \nonumber\\
\end{eqnarray}

Similarly,

\begin{eqnarray}
\label{eq:entropy2}
\int_{\mathbb{R}^{n_{2}}} \int_{\mathbb{R}^{n_{1}}}   f(\mathbf{x_{1}}, \mathbf{x_{2}}) \log \big( f_{2}(\mathbf{x_{2}}) \big)  \mathrm{d} \mathbf{x_{1}} \mathrm{d} \mathbf{x_{2}} \geq  \int_{\mathbb{R}^{n_{2}}} \int_{\mathbb{R}^{n_{1}}}    f(\mathbf{x_{1}}, \mathbf{x_{2}})  \log \big(  h_{2}(\mathbf{x_{2}})  \big) \mathrm{d} \mathbf{x_{1}}\mathrm{d}  \mathbf{x_{2}}.
\end{eqnarray}

Multiplying \eqref{eq:entropy1} and \eqref{eq:entropy2} by -1 and adding them, we have 

\begin{eqnarray*}
\int_{\mathbb{R}^{n_{2}}} \int_{\mathbb{R}^{n_{1}}}     f( \mathbf{x_{1}}, \mathbf{x_{2}}) \log \bigg( \frac{1}{f_{1}(\mathbf{x_{1}})  f_{2}(\mathbf{x_{2}})} \bigg)  \mathrm{d} \mathbf{x_{1}}  \mathrm{d}\mathbf{x_{2}} \leq  \int_{\mathbb{R}^{n_{2}}} \int_{\mathbb{R}^{n_{1}}}   f( \mathbf{x_{1}}, \mathbf{x_{2}}) \log \bigg(\frac{1}{h_{1}(\mathbf{x_{1}}) h_{2}(\mathbf{x_{2}})}  \bigg) \mathrm{d} \mathbf{x_{1}}  \mathrm{d} \mathbf{x_{2}}.
\end{eqnarray*}

Adding $ \displaystyle \int_{\mathbb{R}^{n_{2}}} \int_{\mathbb{R}^{n_{1}}}    f( \mathbf{x_{1}}, \mathbf{x_{2}}) \log \big( f(\mathbf{x_{1}}, \mathbf{x_{2}}) \big)  \mathrm{d} \mathbf{x_{1}} \mathrm{d}  \mathbf{x_{2}} $ to both sides yields the result.

\end{document}